\documentclass[a4paper,11pt]{article}
\usepackage{pos}
\usepackage{cleveref}

\title{Atmospheric muon fluxes at sub-orbital neutrino detectors}
 \ShortTitle{Atmospheric muon fluxes}

\author*[a]{Diksha Garg}
\author[a]{Mary Hall Reno}

\affiliation[a]{University of Iowa,\\
  Iowa City, Iowa, USA}

\emailAdd{diksha-garg@uiowa.edu}
\emailAdd{mary-hall-reno@uiowa.edu}

\abstract{
Very-high-energy and ultra-high-energy neutrinos are messengers of energetic sources in the universe. Sub-orbital and satellite-based neutrino telescopes employ detectors of the atmospheric Cherenkov emission from extensive air showers (EASs) generated by charged particles. These Cherenkov detectors can be pointed below or above the Earth's limb. Cherenkov emissions produced from directions below the limb are from upward-going EASs produced in the atmosphere sourced by Earth-skimming neutrinos. When the Cherenkov telescope is pointed slightly above the Earth's limb, signals from EASs are initiated by cosmic ray interactions in the atmosphere. For sub-orbital detectors, muons produced from cosmic rays in the atmosphere can directly hit the Cherenkov telescope. Using a semi-analytic technique with cascade equations for atmospheric particle fluxes, we quantify the atmospheric muon flux that reaches sub-orbital telescopes like Extreme Universe Space Observatory Super Pressure Balloon 2 (EUSO-SPB2).
We assess this potential background to the EAS signals. The calculation technique may also provide an understanding of the evolution of the muon content in individual EAS.}

\ConferenceLogo{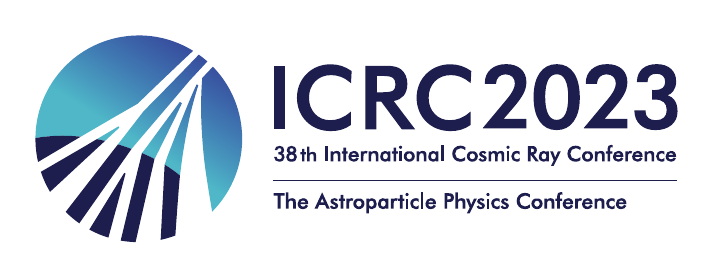}

\FullConference{%
The 38th International Cosmic Ray Conference (ICRC2023)\\
  26 July -- 3 August, 2023\\
  Nagoya, Japan}

\begin{document}
\maketitle

\section{Introduction}
Neutrinos, neutral and weakly interacting particles, can travel astronomical distances unhindered and act as messengers from the distant universe. To detect the very-high-energy (VHE) ($E>10^{15}$ eV) neutrino flux, large volume neutrino targets/detectors are needed. One approach is to use Earth as a neutrino converter, where neutrinos propagating through the Earth interact to produce charged leptons (electrons, muons, and tau-leptons). The charged leptons can exit the Earth and create extensive-air-showers (EASs) in the atmosphere. The EASs have optical Cherenkov, radio emission and fluorescent radiation associated to them. The radiation can be detected by ground-based, sub-orbital, and orbital neutrino telescopes such as IceCube~\cite{IceCube:2021}, EUSO-SPB2~\cite{SPB2,Eser:2021mbp}, and a future POEMMA~\cite{Olinto_2021}. Another feature of cosmic ray interactions in the atmosphere is the generation of atmospheric neutrino and muon fluxes. The atmospheric muon flux incident on EUSO-SPB2 is the subject of this study.

\begin{figure*}[t]
    \centering
    \includegraphics[width=0.65\textwidth]{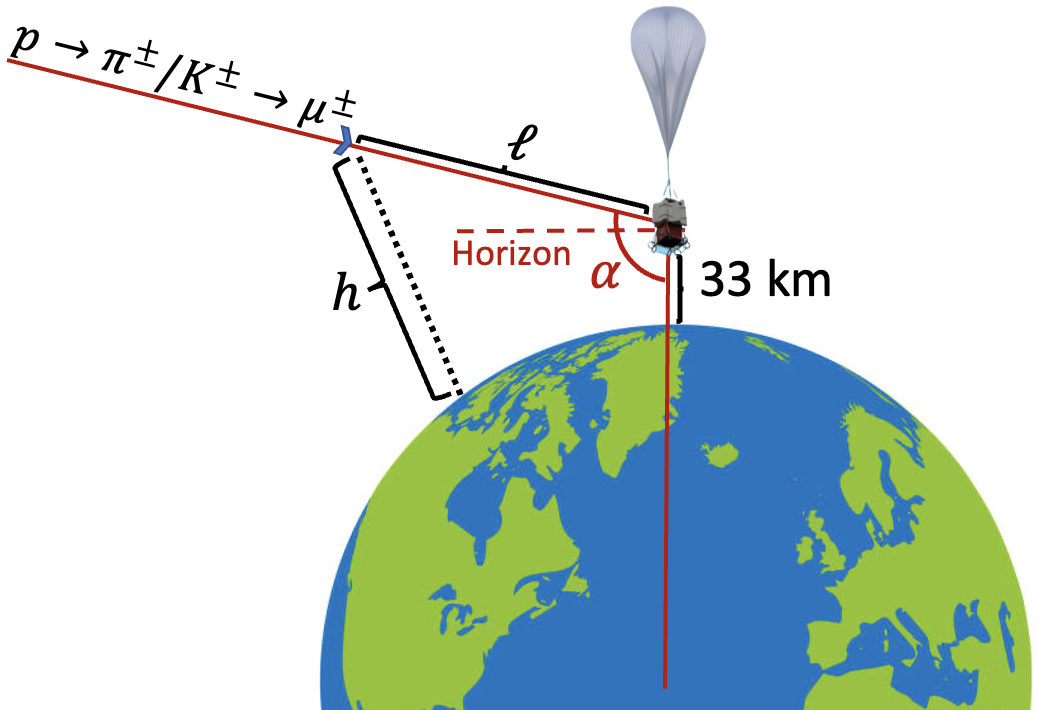}
    \caption{Sub-orbital telescope (like, EUSO-SPB2) at $33$ km altitude. 
    In red is the particle trajectory to the telescope surface at different $\alpha$ angles ($\alpha=0^\circ$ at nadir). The cosmic rays ($p$) interact in the atmosphere to create pions and kaons which decay to muons. At a given point in the particle's trajectory, $h$ represents the altitude of that point and $\ell$ represents the remaining trajectory distance from that point to the telescope. The figure is not to scale. }
    \label{fig:Earth_model}
\end{figure*}

EUSO-SPB2 had a fluorescence telescope (FT) (nadir pointing) and a Cherenkov telescope (CT) that pointed above the limb to detect EAS from cosmic rays. It flew at an altitude of 33 km from the surface of the Earth. 
The muons produced from cosmic rays interactions in the atmosphere produce charged pions and kaons that can decay to muons. These muons can directly hit the CT and the FT. 
This is depicted in~\cref{fig:Earth_model}, where $\alpha$ describes the incident direction of  the cosmic rays,
which can be below or above the telescope's horizon. For the telescopes at an altitude of 33 km, $\alpha>84.2^\circ$ for cosmic ray trajectories to be above the Earth's limb. 

Direct muon hits from the atmospheric muon flux can act as a potential background to the EAS signals measured by the CT and the FT. The atmospheric lepton fluxes are well-studied for ground-based instruments (see, e.g., refs. \cite{Lipari:1993hd,Gaisser:2019xlw,Gaisser:2002jj,Fedynitch:2018cbl,Kozynets:2023tsv}), but less so for sub-orbital instruments like EUSO-SPB2.
Here, we make a first estimate of the rate of atmospheric muons hitting the EUSO-SPB2 telescopes.

\section{Atmospheric lepton fluxes with the $Z$-moment approximation}\label{sec:zmom_approx}

The atmospheric particle flux $\phi_j(E,X)$ for particle $j$, as a function of energy $E$ and column depth $X$ can be written as 
\begin{eqnarray}
\frac{d\phi_j(E,X)}{dX}&=&-\frac{\phi_j(E,X)}{\lambda_j(E)} - \frac{\phi_j(E,X)}{\lambda_j^{\rm dec}(E)}
+ \sum S(k\to j)\, ,\\
\label{eq:source_term} S(k\to j) &=& \int_E^{\infty}dE ' \frac{\phi_k(E',X)}{\lambda_k(E')}
\frac{dn(k\to j;E',E)}{dE}\, ,
\end{eqnarray}
for all particles except for muons.  The symbol $\lambda_j(E)$ and $\lambda_j^{\rm dec}(E)= E \tau_{j}\rho/m_j$ are the interaction length and the decay length, respectively, of the particle $j$ in the atmosphere. The quantities $E$, $\tau_j$, $m_j$ are the energy, lifetime and mass of the particle $j$. We have suppressed the angular ($\alpha$) dependence of $\phi_j$ and $X$. The column depth is given as:
\begin{equation}\label{eq:col_dpeth}
    X(\ell,\alpha) = \int_\ell^\infty d\ell' \rho(h(\ell',\alpha)),
\end{equation}
where $h(\ell,\alpha)$ is the altitude, $\ell$ is the trajectory distance and $\alpha$ is the nadir angle, also shown in~\cref{fig:Earth_model}. The atmospheric density ($\rho$) considered here is given by an exponential distribution, 
\begin{equation}\label{eq:rho_atm}
\rho = \rho_0\exp(-h/h_0)\ ,
\end{equation}
where $h_0=6.4$ km and $\rho_0 h_0=1300$ g/cm$^2$. We approximate particle trajectories as one-dimensional, a good approximation above 10 GeV \cite{Kozynets:2023tsv}. The interaction and decay distributions in the source term $S(k\to j)$ (eq.~\eqref{eq:source_term}) in terms of cross-section ($\sigma$) and decay width ($\Gamma$) are
\begin{eqnarray}
\frac{dn(k\to j;E',E)}{dE}&=& \frac{1}{\sigma_{kA}(E')}\frac{d\sigma(kA\to jY;E',E)}{dE }
\quad {\rm (interaction)\, ,}\\
\frac{dn(k\to j;E',E)}{dE}&=& \frac{1}{\Gamma_k(E')}\frac{d\Gamma (k\to jY;E',E)}{dE}\quad {\rm (decay)}
\ .
\end{eqnarray}
For muons, electromagnetic energy loss is incorporated in the continuous-energy-loss approximation, where
\begin{equation}
    \Biggl\langle \frac{dE}{dX}\Biggr\rangle \simeq \frac{dE}{dX}\simeq -(a+bE)\equiv -\beta(E)\,.
\end{equation}
Here, $a$ and $b$ are the energy loss parameters accounting for ionization, and bremsstrahlung, pair production and photo-nuclear energy loss, respectively. In this approximation
\begin{equation}
    \frac{d\phi_\mu(E,X)}{dX}= - \frac{\phi_\mu(E,X)}{\lambda_j^{\rm dec}(E)}
    +\frac{\partial}{\partial E}\Bigl[ \beta(E)\phi_\mu(E,X)\Bigr]
+ \sum S(k\to \mu)\,. 
\end{equation}

We use the same analytic approximation method involving spectrum-weighted $Z$-moments that is successful for calculating the atmospheric lepton fluxes at the Earth's surface~\cite{Lipari:1993hd,Gaisser:2019xlw}.
The analytic approximation to determine atmospheric lepton fluxes relies on flux-weighted integrals of differential distributions for particle production and decay. The source of particle $j$ from initial particle $k$, denoted $S(k\to j)$, is given by eq.~\eqref{eq:source_term} and can be re-written in terms of $Z$-moment as 
\begin{eqnarray}
\nonumber
S(k\to j)
&\simeq & 
\Biggl[\int _E^{\infty}dE ' \frac{\phi_k^0(E')}{\phi_k^0(E)}
\frac{\lambda_k(E)}{\lambda_k(E')}
\frac{dn(k\to j;E',E)}{dE}\Biggr] \frac{\phi_k(E,X)}{\lambda_k(E)}\\
&\equiv&Z_{kj}(E)  \frac{dn(k\to j;E',E)}{dE}\, ,
\label{eq:source_Z}
\end{eqnarray}
where $\phi_k (E,X)=\phi_k^0(E)f(X)$ and
 $\Lambda_k=\lambda_k/(1-Z_{kk})$ so that
$f(X) = \exp{(-X/\Lambda_k)}$. 

\begin{figure*}[t]
    \centering
    \includegraphics[width=0.49\textwidth]{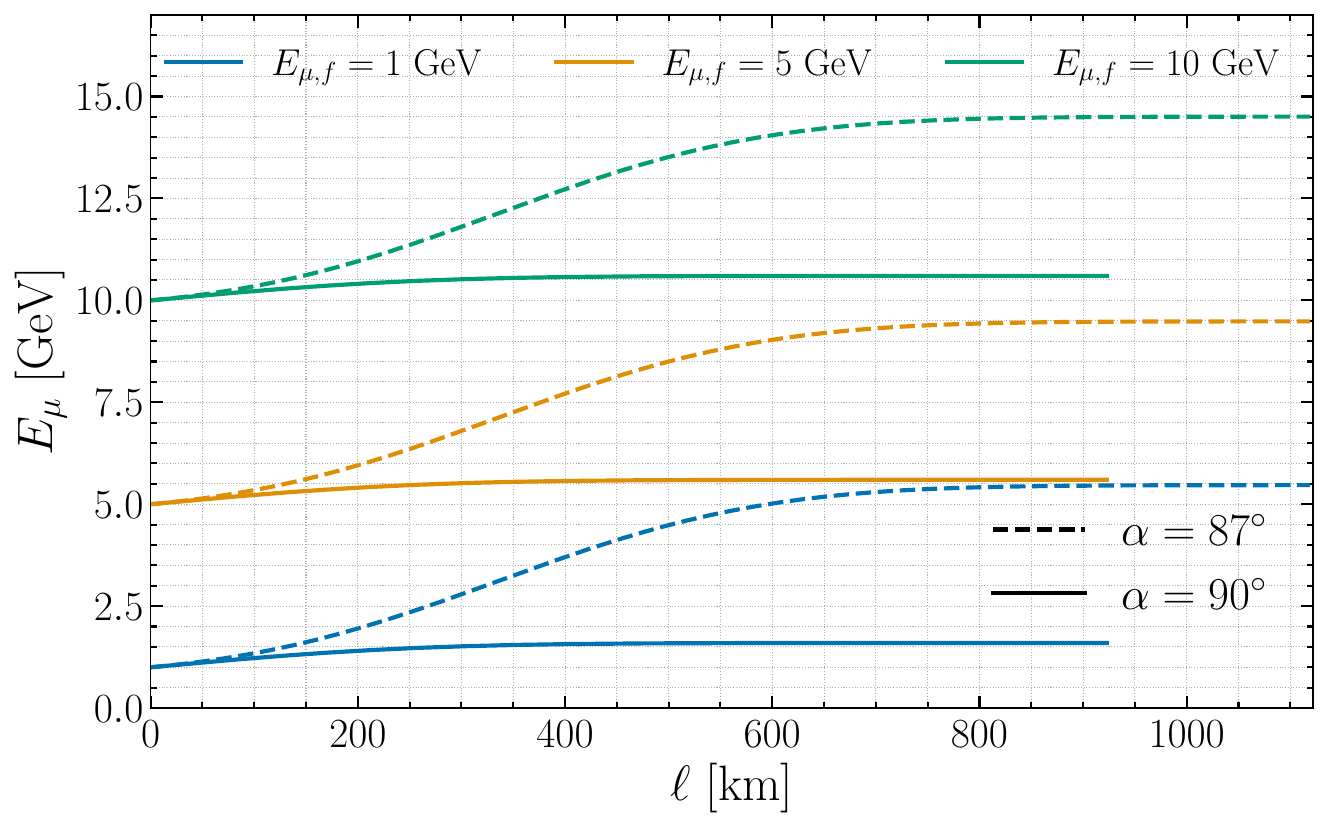}
    \includegraphics[width=0.49\textwidth]{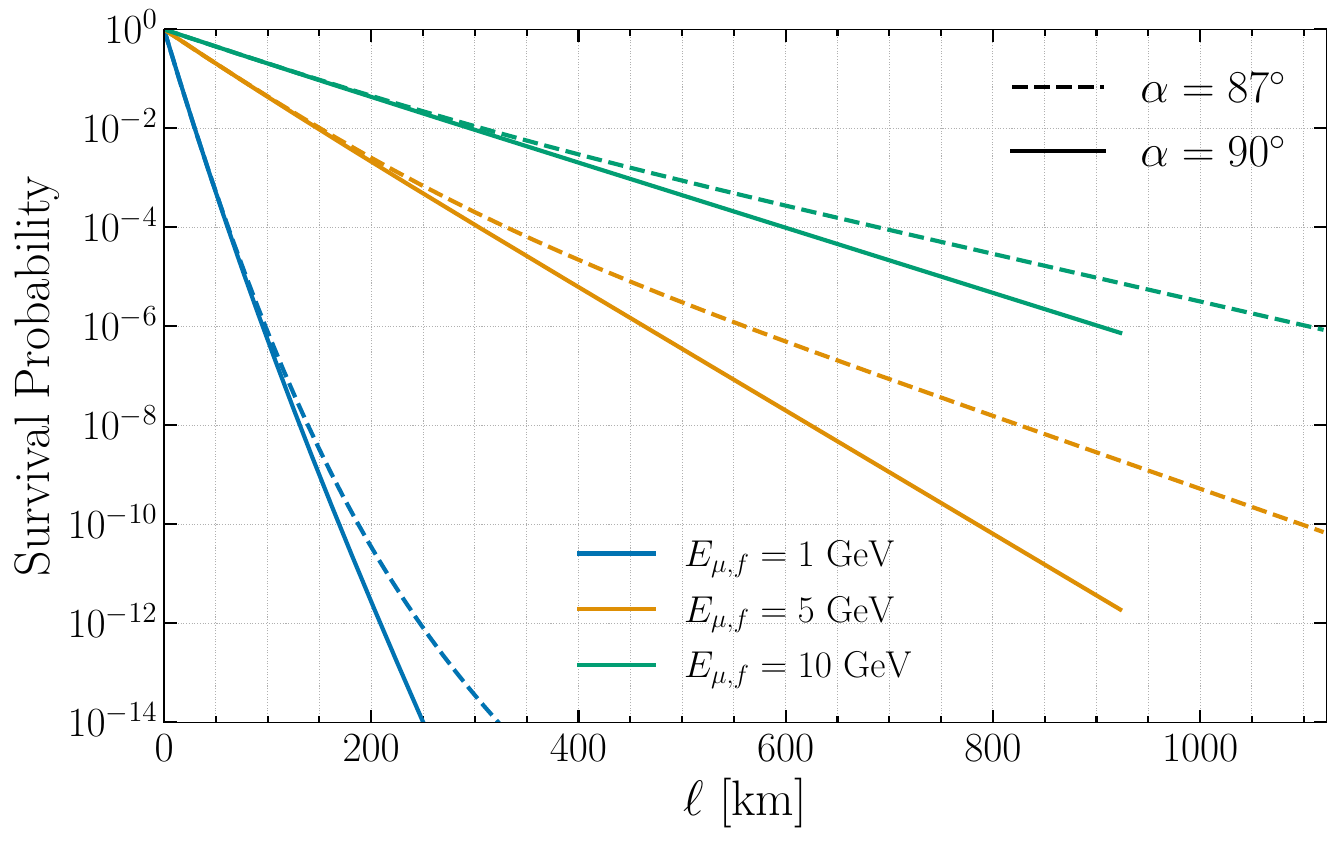}
    \caption{Muon energy (left) and the survival probability of muon (right) produced $\ell$ distance away from the EUSO-SPB2 balloon. The balloon is at $\ell=0$ km. It is shown for three final muon energies at the balloon. The dashed line is for $\alpha=87^\circ$ and solid line is for $\alpha=90^\circ$. The plot on the left shows the muon reaching the balloon (at $\ell=0$ km) with a final energy of $E_{\mu,f}=1,5,10$ GeV starts its trajectory at a higher energy.}
    \label{fig:energy_loss_surv_prob}
\end{figure*}
To evaluate the flux of muons, we step through the column depth of the atmosphere with steps $\Delta X$ that are small relative to $\lambda_\mu^{\rm dec}$ and $\lambda_j$ for interactions and decays.  For cosmic ray nucleons $N$, and for the pions and kaons they produce in the atmosphere,  this translates for each step to
\begin{eqnarray}\label{eq:phi_CR}
    \phi_N(E,X+\Delta X) &=& \phi_N(E, X)\Biggl(1-\frac{\Delta X}{\Lambda_N}\Biggr)\\
    \label{eq:phi_j}\phi_j(E,X+\Delta X) &=& \phi_j(E, X)\Biggl(1-\frac{\Delta X}{\Lambda_j}
    -\frac{\Delta X}{\lambda_j^{\rm dec}}\Biggr)
    + Z_{N\to j}\phi_N(E,X)\frac{\Delta X}{\lambda_N}   
\end{eqnarray}
where $j=\pi,K$. For muons
\begin{eqnarray}
\label{eq:phimu}
    \phi_\mu(E,X+\Delta X) &=& \Biggl[\phi_\mu(E', X)\Biggl(1
    -\frac{\Delta X}{\lambda_\mu^{\rm dec}}\Biggr)
    + \sum_{j=\pi,K} Z_{j\to \mu}\phi_j(E',X)\frac{\Delta X}{\lambda_j^{\rm dec}}\Biggr] \exp(b\Delta X) \\
    \label{eq:energyi}
    E' &=& (E+a/b)\exp(-b\Delta X) -a/b\,.
\end{eqnarray}
To simplify the calculation further, we make several approximations. 
The energy loss parameters $a$, $b$, and the $Z$-moments are taken to be energy independent with their respective values are taken from ref.~\cite{Lipari:1993hd}, including pion and kaon decays as sources of muons in the atmosphere. We don't account for the Earth's magnetic field, an effect most important for muons with energies below 10 GeV \cite{Kozynets:2023tsv}. 
We approximate the cosmic ray flux ($\phi_N^0$) as a function of energy per nucleon $E$ by 
\begin{equation}
\nonumber
\phi_N^0(E) \Biggl[\frac{\rm nucleons}{\rm cm^2\, s\, sr\,
  GeV}\Biggr]
= 1.7\ (E/{\rm GeV})^{-2.7},\quad E< 5\cdot 10^6\ {\rm GeV}\ .
\end{equation}

The muon flux is evaluated according to eq.~\eqref{eq:phimu}. Our evaluation of the atmospheric muon flux on the ground is in agreement with the results in ref. \cite{Lipari:1993hd}. For total atmospheric column depth $X(\alpha)$, the muon energy at the beginning of the trajectory has initial energy
$E_i=(E+a/b)\exp(bX)-a/b$, and for each of the $n$ steps in column depth, the muon energy is decreased according to eq.~\eqref{eq:energyi}. Therefore, the muons reaching the telescope with a final energy $E_{\mu,f}$ ($E_\mu=E_{\mu,f}$ at $\ell=0$) will have started their journey with a higher initial energy, as shown in the left panel of \cref{fig:energy_loss_surv_prob} for two different column depths corresponding to $\alpha = 87^\circ$ and $\alpha=90^\circ$.  Muon energy loss also impacts the survival probability of the muons, as  shown in the left panel of \cref{fig:energy_loss_surv_prob} which shows the muon survival probability as a function of $\ell$ for three final muon energies and two column depths.

\section{Results}

\begin{figure*}[t]
    \centering
    \includegraphics[width=0.49\textwidth]{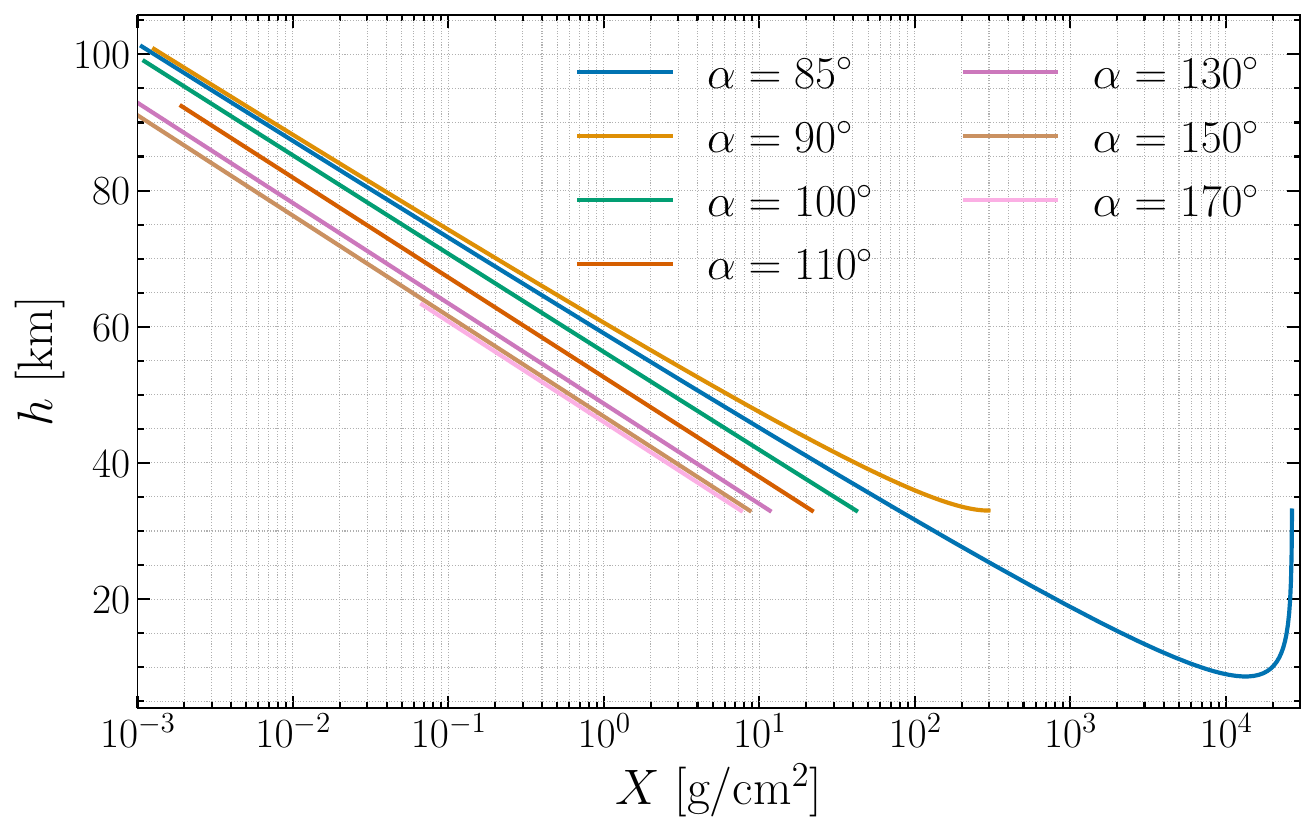}
    \includegraphics[width=0.49\textwidth]{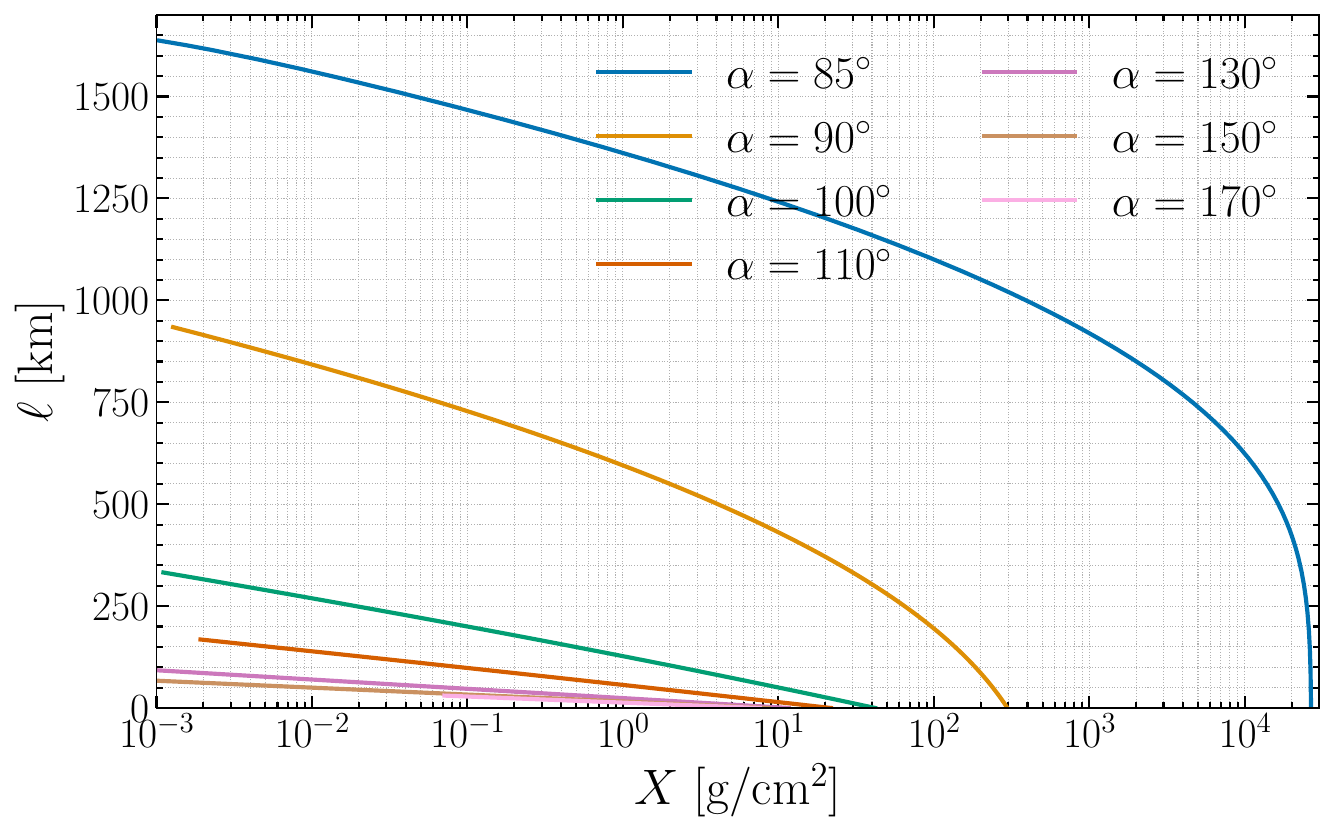}
    \caption{Left: Altitude $h$ (left) and trajectory distance $\ell$ (right) as a function of column depth $X$ starting from very high altitude, for trajectories below the horizon, at the horizon and above the horizon (different $\alpha$ values) for an instrument at an altitude of 33 km. The atmospheric density is approximated by an exponential function. 
    }
    \label{fig:col_alt_below}
\end{figure*}

\begin{figure*}[t]
    \centering
    \includegraphics[width=0.49\textwidth]{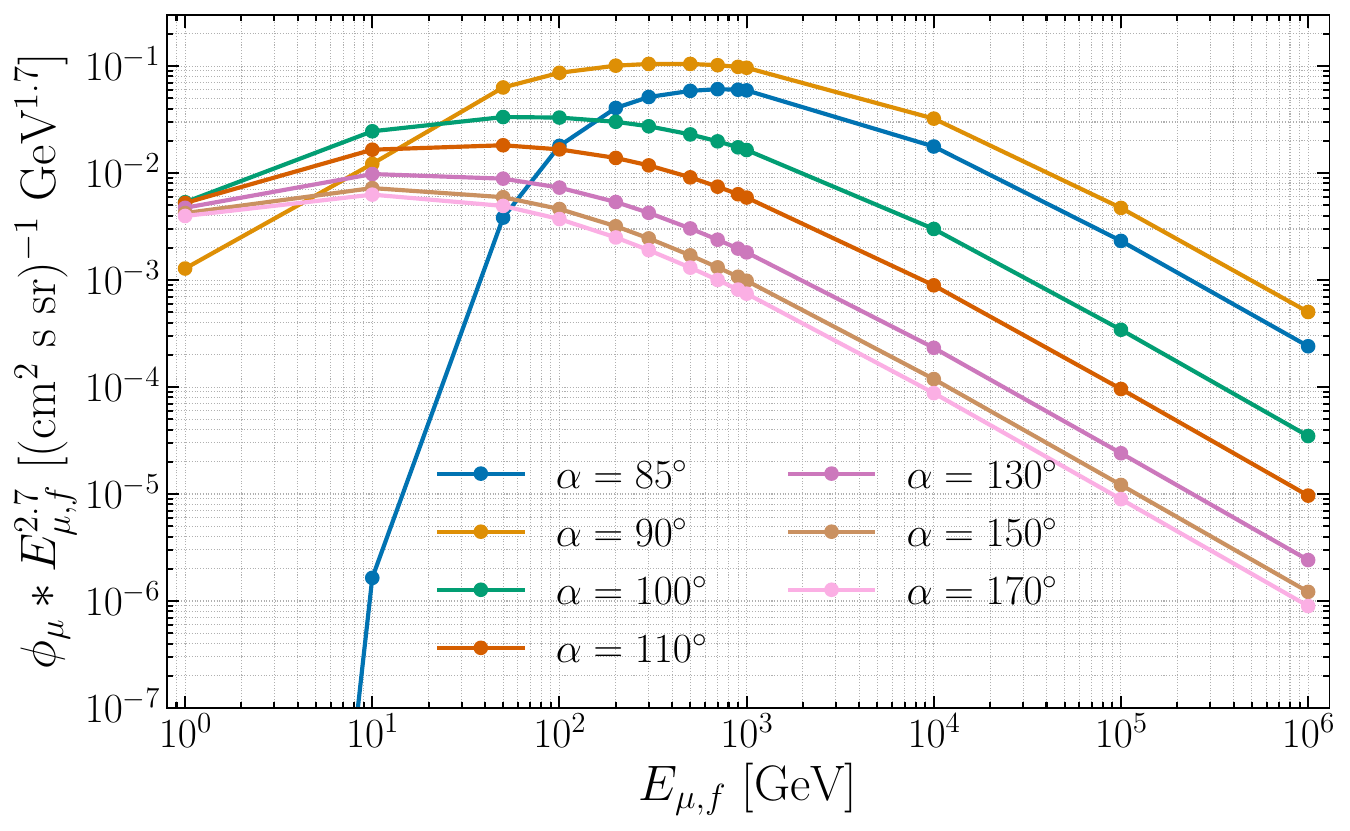}
    \includegraphics[width=0.49\textwidth]{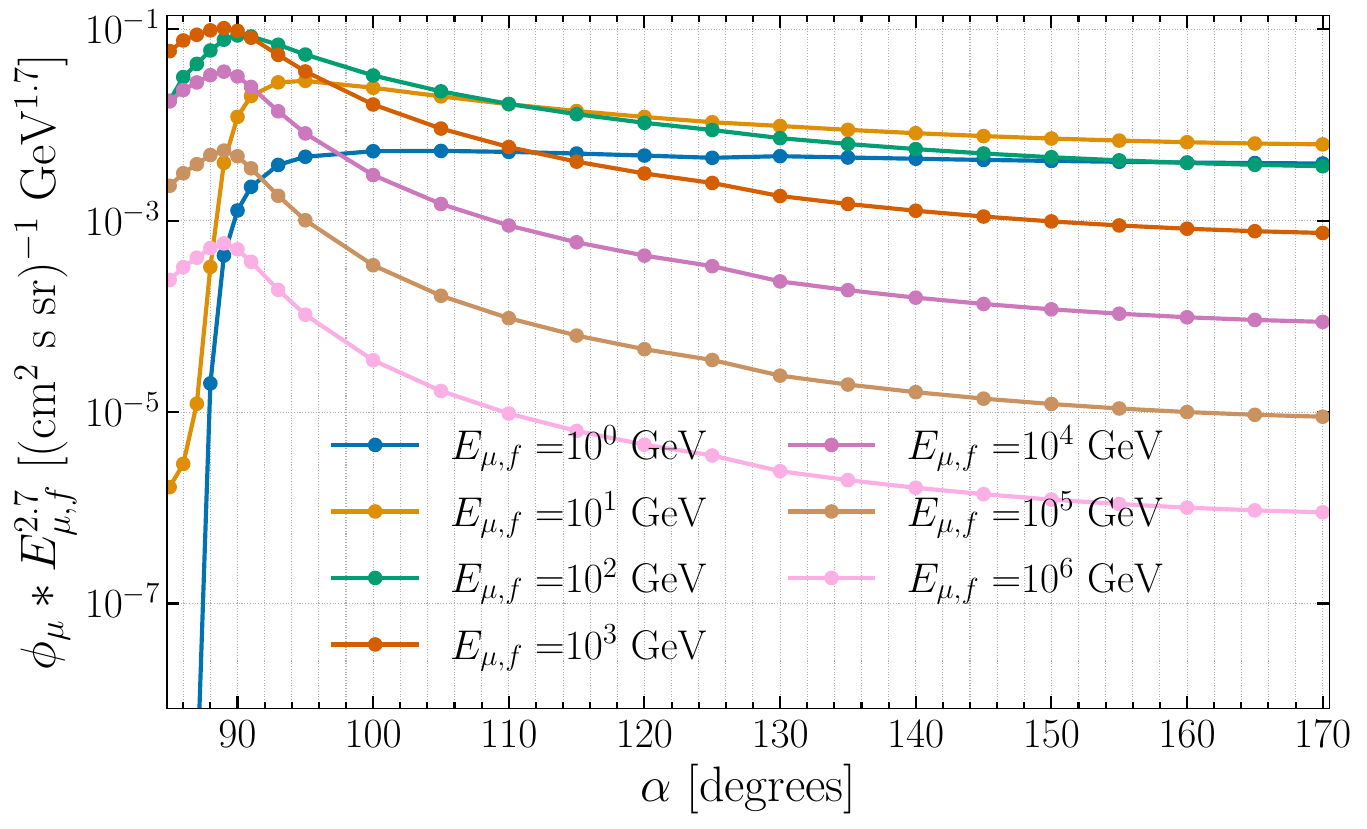}    
    \caption{Muon flux (scaled by $E_{\mu,f}^{2.7}$) as a function of final muon energy reaching the detector. It is shown for different trajectories above and below the horizon. Right: Muon flux (scaled by $E_{\mu,f}^{2.7}$) as a function of $\alpha$. It is shown for different final muon energies reaching the detector.}
    \label{fig:flux_results}
\end{figure*}

\begin{figure*}[t]
    \centering
    \includegraphics[width=0.6\textwidth]{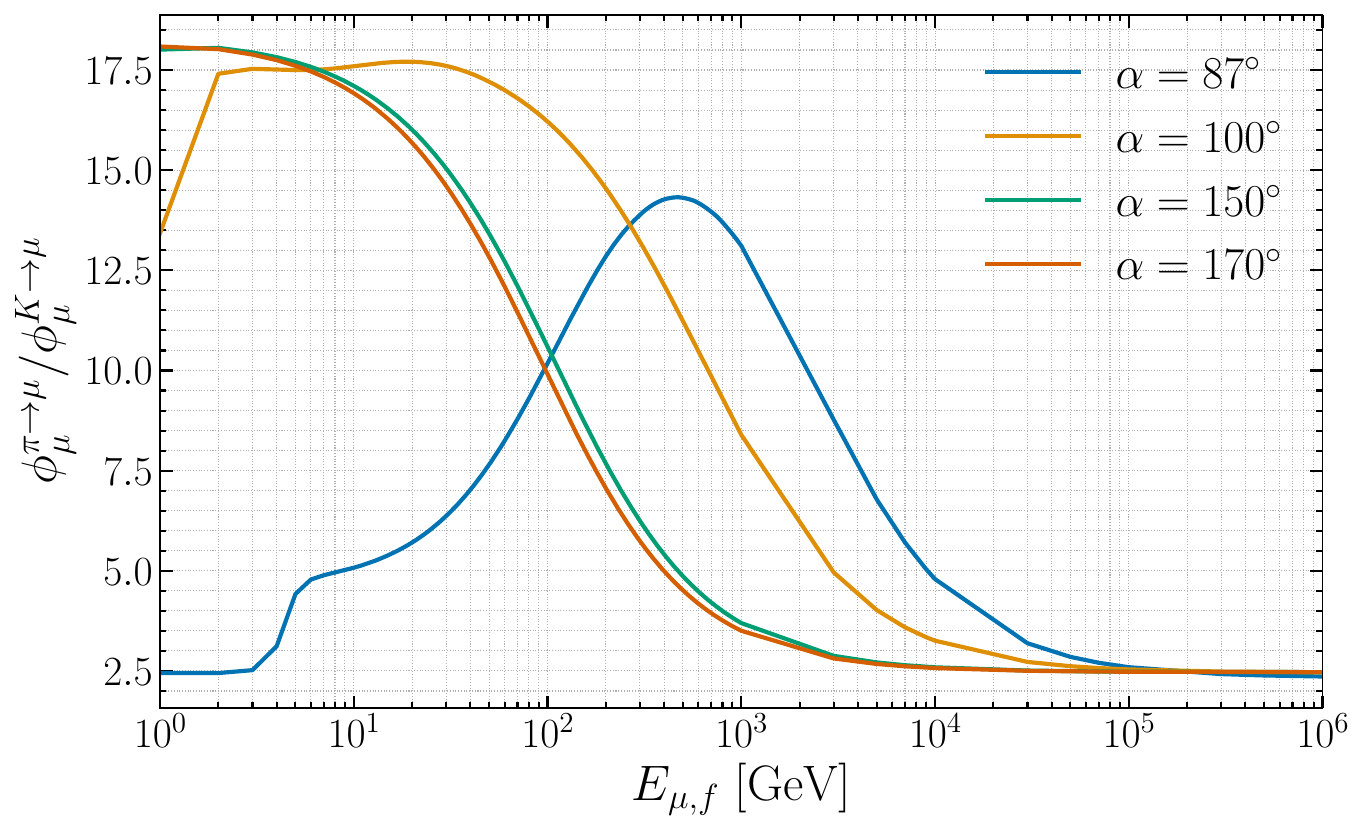}
    \caption{The ratio of muon flux produced from charged pions to charged kaons, as a function of final muon energies. It is shown for different trajectories below and above the horizon.  }
    \label{fig:pion_kaon_flux}
\end{figure*}

\begin{figure*}[t]
    \centering
    \includegraphics[width=0.65\textwidth]{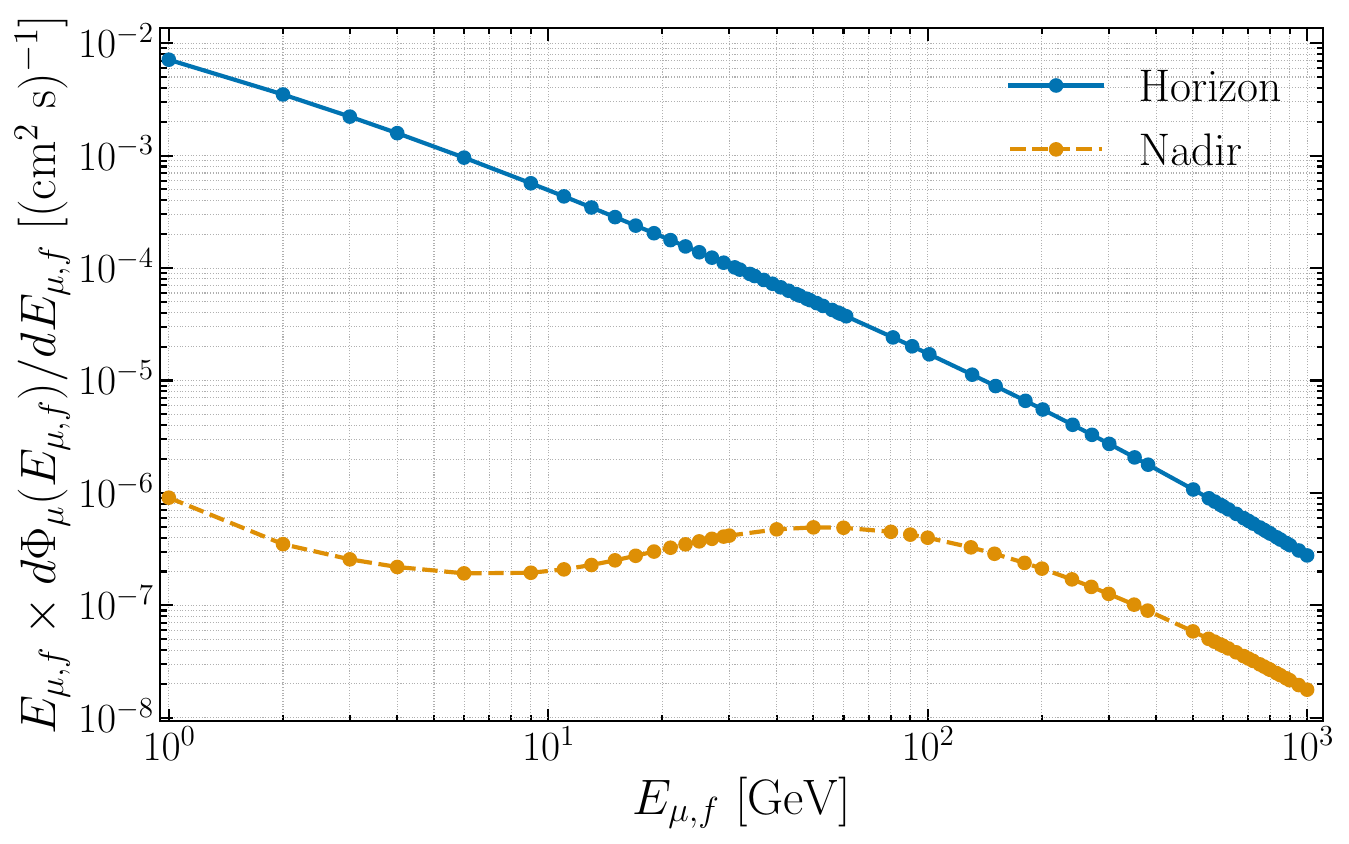}
    \caption{Muon flux reaching per cm$^2$ of area of the detector per second scaled by final muon energies, as a function of final muon energy. Solid line is for the detector pointing to the horizon, and dashed line is for the detector pointing in nadir direction. }
    \label{fig:muon_number}
\end{figure*}

The cosmic rays interact after traversing $X \simeq 120$ g/cm$^2$ in the atmosphere to form charged mesons, like pions and kaons, which can decay to muons. The column depth depends on the density of the air in the atmosphere, and is given by eq.~\eqref{eq:col_dpeth}. The atmosphere near the surface of the Earth is denser and its density decreases as we go farther up from the surface. Therefore, understanding the column depth variation along a particle's trajectory as a function of $h$ and $\ell$ is important.~\Cref{fig:col_alt_below} shows the column depth variation in the atmosphere as a function of altitude (left) and particle's trajectory distance (right). It is shown for different trajectories above and below the horizon. 

Our main results for the muon flux with energy $E_{\mu,f}$ at a detector at an altitude of 33 km are shown in fig. \ref{fig:flux_results}.
~\Cref{fig:flux_results} (left) shows the muon flux scaled by the $E_{\mu,f}^{2.7}$, as a function of $E_{\mu,f}$, for several values of $\alpha$, from both below and above the horizon of the telescope.~\Cref{fig:flux_results} (right) shows the muon flux scaled by $E_{\mu,f}^{2.7}$ as a function of $\alpha$ for fixed final muon energies at the balloon. 

The particle trajectories with smaller $\alpha$, have to travel longer distances to reach the balloon as compared to the particle trajectories with higher $\alpha$. For higher $\alpha$, the density of the atmosphere is lower, and it leads to the cosmic rays interacting to produce charged pions and kaons closer to the detector. This means the production of muons is also closer to the detector, increasing the muon flux reaching the telescopes on the balloon for higher $\alpha$ trajectories. This is why the muon flux reaching the balloon with a final muon energy of 1 GeV, for example, is lower for smaller $\alpha$ but increases with higher $\alpha$ (shown in~\cref{fig:flux_results} right). 

We can see in~\cref{fig:flux_results} (left), the muon flux increases with increasing muon final energies as more muons are able to reach the balloon. But the flux starts falling for $E_{\mu,f} \gtrsim 10^3$ GeV. This is because fewer muons at higher energies are being produced as the probability of interaction for pions and kaons is higher than their decay probability.  

For reference, we show the ratio of the muon flux reaching the balloon produced from charged pion decays to the flux from charged kaon decays in~\cref{fig:pion_kaon_flux} for different values of $\alpha$. The muon flux from pion decays is a factor of $\sim 8-11$ higher than the flux from kaon decays at lower final muon energies. For higher final muon energies, the ratio of the muon flux produced from pions to kaons is nearly constant as a function of energy, with a value of $\sim 1.5$.  

To calculate the rate of muons incident on the CT pointed horizontally, and from the upward muon flux incident on the FT pointing to the nadir, we integrate the flux of muons over the appropriate solid angle, starting with $\alpha=84.2^\circ$ at the Earth's limb.
For these two cases, the flux integrated over solid angle and energy for energy $E_{\mu,f}$, denoted by $\Phi_\mu(E_{\mu,f})$ are
\begin{eqnarray}
    \Phi_\mu(E_{\mu,f}) &\equiv& \int_{E_ {\mu,f}} dE_\mu \int_{\varphi=-\pi/2}^{\varphi=\pi/2}\int_{\alpha=84.2^\circ}^{\alpha=180^\circ}\phi_\mu(E_{\mu},\alpha) \sin^2\alpha \cos\varphi \, d\alpha 
    d\varphi, \quad {\rm horizontal}\\
    \Phi_\mu(E_{\mu,f}) &\equiv& \int_{E_ {\mu,f}} dE_\mu \int_{\varphi=0}^{\varphi=2\pi}\int_{\alpha=84.2^\circ}^{\alpha=90^\circ}\phi_\mu(E_{\mu},\alpha) \cos\alpha \sin\alpha\, d\alpha 
    d\varphi, \quad {\rm nadir}\,.
\end{eqnarray}

The results for $E_{\mu,f}\times d\Phi_\mu/dE_{\mu,f}$ are shown in~\cref{fig:muon_number} as a function of final muon energy are shown for each orientation. The rate of up-going muons in the atmospheric flux that reach the FT, parallel to the Earth (pointing to the nadir) is much smaller than the rate incident on the 
CT pointing to the horizon.
Integrating over energy with a minimum energy of $E_{\mu,f}=1$ GeV yields $6.6\times 10^{-3}$/cm$^2$/s on the horizontally-pointing CT, and $2.1\times 10^{-6}$/cm$^2$/s on the nadir-pointing FT. The flux of muons on the ground is  approximately 1/cm$^2$/min$\,\simeq 1.7\times 10^{-2}$/cm$^2$/s~\cite{PDBook}.

While our approximation that for a given $\alpha$ we can use the cascade equations in a 1-dimensional approximation is good for $E_{\mu,f}\gtrsim 10$ GeV, is not as good an approximation for lower energy muons where the Earth's magnetic field bends the muon trajectory (and makes it longer) \cite{Fedynitch:2022vty}. For reference, $\Phi_\mu(E_{\mu,f}=10$ GeV) is 
$3.6\times 10^{-4}$/cm$^2$/s for the horizontal orientation of the telescope, and $1.3\times 10^{-6}$/cm$^2$/s for the nadir orientation. 

\section{Discussion}

Our muon flux calculation is an approximate result, but for the orientation of CT pointing to its horizon, or near to it, is interesting enough to pursue with a more detailed evaluation. As muons pass through the plane of the detector, they are ionizing and deposit energy. More detailed modeling of the detector is required to understand their impact. 

Given the areas of the CT, if we assume that 100\% of the muon rate triggers the detector, the atmospheric muon rate for muons above 1 GeV is 1.21 Hz and above 10 GeV is 0.066 Hz, on the CT of area  184 cm$^2$ with horizontal pointing. There may be an opportunity to measure the atmospheric muon flux with EUSO-SPB2 using data collected to determine the air glow background. 

\vspace{0.2cm}
\noindent{\bf Acknowledgements}
We thank J. Krizmanic and J. Szabelski for illuminating discussions. This work was supported in part the the US Department of Energy grant DE-SC-0010113.

\bibliographystyle{JHEP}
\bibliography{references}

%
%
%

\end{document}